\documentclass[twoside,a4paper,11pt]{sea10}
\usepackage{graphicx}
\usepackage{hyperref}
\usepackage{movie15}
\begin{document}
\pagenumbering{arabic}
\setcounter{page}{187}
\pagestyle{myheadings}
\thispagestyle{plain}
{\flushleft\includegraphics[width=\textwidth,bb=58 650 590 680]{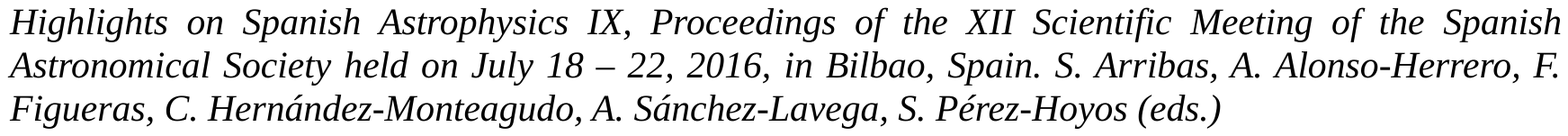}}
\vspace*{0.2cm}
\begin{flushleft}
{\bf {\LARGE
%
Relic galaxies: where are they?
%
}\\
\vspace*{1cm}
%
Luis Peralta de Arriba$^{1,2,3}$,
Vicent Quilis$^{4,5}$,
Ignacio Trujillo$^{3,2}$
Mar\'{\i}a Cebri\'an$^{3,2}$
and
Marc Balcells$^{1,3,2}$
%
}\\
\vspace*{0.5cm}
%
$^{1}$
Isaac Newton Group of Telescopes, E-38700 Santa Cruz de La Palma, La Palma, Spain\\
$^{2}$
Universidad de La Laguna, Departamento de Astrof\'{\i}sica, E-38206 La Laguna, Tenerife, Spain\\
$^{3}$
Instituto de Astrof\'{\i}sica de Canarias (IAC), E-38205 La Laguna, Tenerife, Spain\\
$^{4}$
Departament d'Astronomia i Astrof\'{\i}sica, Universitat de Val\`encia, E-46100 Burjassot, Val\`encia, Spain\\
$^{5}$
Observatori Astron\`omic, Universitat de Val\`encia, E-46980 Paterna, Val\`encia, Spain
%
\end{flushleft}
%
\markboth{
Relic galaxies: where are they?
}{ 
%
Peralta de Arriba et al.
%
}
\thispagestyle{plain}
\vspace*{0.4cm}
\begin{minipage}[l]{0.09\textwidth}
\ 
\end{minipage}
\begin{minipage}[r]{0.9\textwidth}
\vspace{1cm}
\section*{Abstract}{\small
%
The finding that massive galaxies grow with cosmic time fired the starting gun for the search of objects which could have survived up to the present day without suffering substantial changes (neither in their structures, neither in their stellar populations).

Nevertheless, and despite the community efforts, up to now only one firm candidate to be considered one of these relics is known: NGC~1277. Curiously, this galaxy is located at the centre of one of the most rich near galaxy clusters: Perseus. Is its location a matter of chance? Should relic hunters focus their search on galaxy clusters?

In order to reply this question, we have performed a simultaneous and analogous analysis using simulations (Millennium I-WMAP7) and observations (New York University Value-Added Galaxy Catalogue). Our results in both frameworks agree: it is more probable to find relics in high density environments.
%
\normalsize}
\end{minipage}
%
%
%
\section{Scientific motivation}

Massive galaxies are on average more compact in the primitive Universe (e.g. \cite{2008ApJ...687L..61B}, \cite{2007MNRAS.382..109T}). If the mechanism which is growing in size these galaxies is stochastic (such as mergers), there should exist relic galaxies (that is, survivors to this size growth). The discovery of relic galaxies is interesting because they are ideal tools for studying galaxy formation, as they remain intact the imprint of their primal properties. Unfortunately, by now only one firm candidate has been reported (\cite{2014ApJ...780L..20T}). Consequently, the question about which is the best environment for looking for massive relic galaxies arises as an interesting topic. We addressed this question in our recent article \cite{2016MNRAS.461..156P}, which we will summarize on what follows.

\section{Previous works about the preferred environment of relic galaxies}

\begin{itemize}
\item \cite{2013ApJ...762...77P} reported that the fraction of old superdense galaxies is higher in clusters.
\item \cite{2016MNRAS.461..156P} reported that NGC~1277 is a firm candidate to be a relic galaxy. Interestingly, NGC~1277 inhabits the central region of the Perseus galaxy cluster.
\item Using the BOLSOI simulation, \cite{2015MNRAS.449.2396S} reported that the probability of finding dark matter haloes like that of NGC~1277 increases significantly with the mass of the host structure.
\end{itemize}

\section{Our work}

In our recent article \cite{2016MNRAS.461..156P}, we address the question about which is the best environment for looking for massive relic galaxies. We consider as massive galaxies those with stellar masses above $10^{10} \ \mathrm{M_\odot}$. We performed our analysis simultaneously and analogously using simulations and observations. In particular, we used the simulations from Millenium I-WMAP7 (\cite{2013MNRAS.428.1351G}, which contains 1850648 massive galaxies) and the observational information from New York University Value-Added Galaxy Catalogue (\cite{2005AJ....129.2562B}, which contains 41716 massive galaxies).

We define the environmental density around each galaxy in the simulations and observations in the same way: for each one we identify all its massive neighbours within a sphere of radius $R = 0.5~\mathrm{Mpc}$, and compute the following formula:
\begin{equation} \label{eq:rho}
\rho_i = \frac{1}{\frac{4}{3} \pi R^3} \sum_{k=1}^{N_i} M_{i,k},
\end{equation}
where $M_{i,k}$ is the stellar mass of the $k$th neighbour located at less than $R$ of the $i$th galaxy of the sample (galaxy $i$ has $N_i$ neighbours). It is worth noting that, although the prescription is the same for both frameworks, observational computations will be affected by the fingers-of-god effect (and we will emphasize this difference noting that environmental density by $\tilde{\rho}$).

\subsection{Simulations}

\subsubsection{Definition of a relic galaxy in the simulations}

\begin{enumerate}
\item The galaxy must be already formed at $z \sim 2$.
\item The stellar mass of the $z \sim 2$ precursor is higher than the 90 per cent of the $z \sim 0$ mass.
\end{enumerate}

\subsubsection{Results from the simulations}

\begin{figure}
\center
\includegraphics{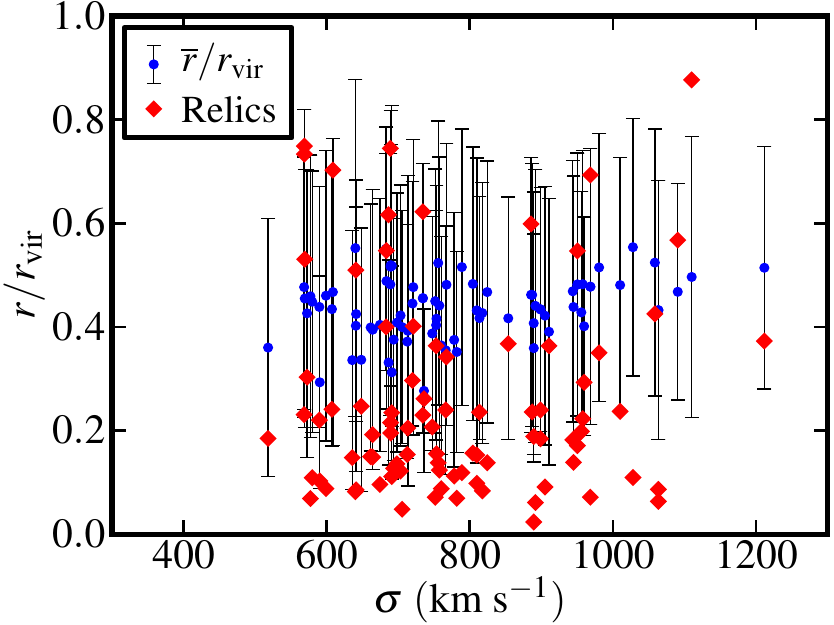}
\caption{\label{fig1}\, From the numerical catalogue, normalized radial position of relic galaxies in clusters (red diamonds) as a function of the velocity dispersion of the clusters, compared with the averaged radial positions $\overline{r}$ of all galaxies in each cluster (blue circles). Error bars represent one standard deviation for the radial distribution of the galaxies in a given cluster. Figure from \cite{2016MNRAS.461..156P}.}
\end{figure}

\begin{figure}
\center
\includegraphics{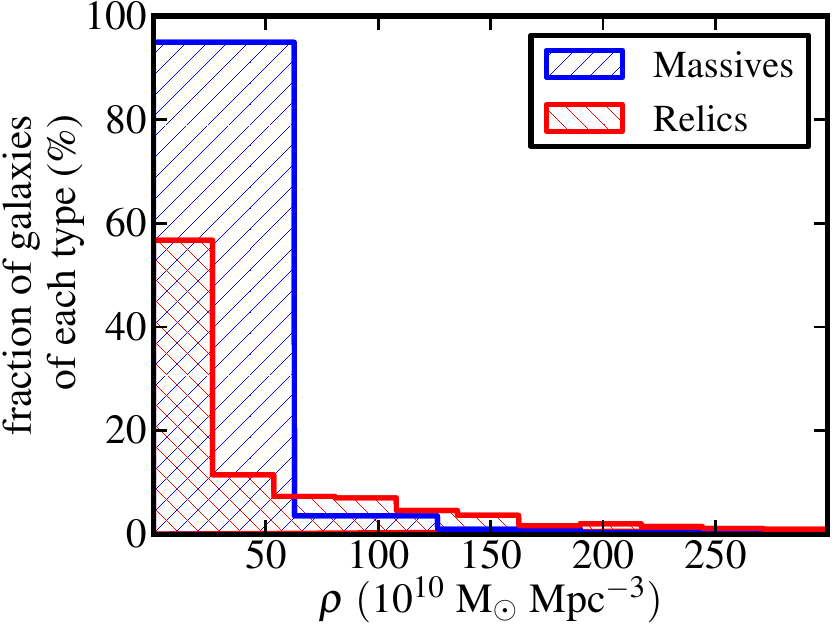}
\caption{\label{fig2}\, From the numerical catalogue, distribution of the environmental density of massive (blue) and relic (red) galaxies. Figure from \cite{2016MNRAS.461..156P}.}
\end{figure}

\begin{itemize}
\item 11.7 per cent of relic galaxies are located in clusters, while 2.6 per cent of massive galaxies live in clusters. This means that the fraction of relic galaxies is 4.6 higher in cluster environments.
\item The relics tend to be located a factor of 2 closer to the central parts of the clusters than the other massive galaxies (Fig.~\ref{fig1}).
\item The fraction of relic galaxies is higher in environments with higher environmental densities (c.f. Fig.~\ref{fig2}).
\end{itemize}

\subsection{Observations}

\subsubsection{Definition of a relic galaxy in the observations}

\begin{figure}
\center
\includegraphics{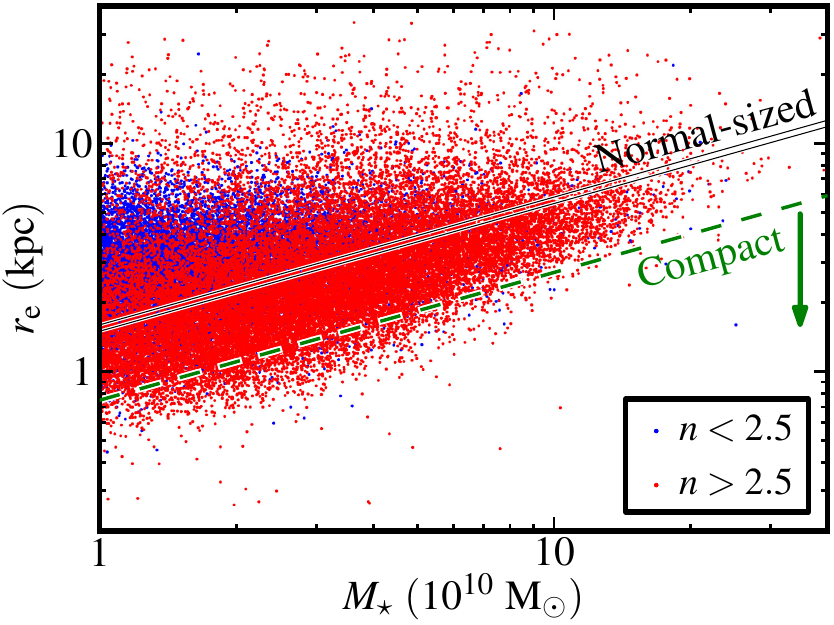}
\caption{\label{fig3}\, From the observational catalogue, distribution over the stellar mass--size plane of the galaxies. The green dashed line indicates the upper limit of the compact definition, while the two solid black lines show the region populated by the galaxies defined as normal-sized. In addition, we have represented the galaxies with different colours depending on their S\'ersic indices as indicated in the legend. Figure from \cite{2016MNRAS.461..156P}.}
\end{figure}

\begin{enumerate}
\item Its stellar population is old: its position in the rest-frame colour--colour diagram $g-r$ versus $r-i$ corresponds to the location of a single stellar population with an age older than 10 Gyr. 
\item Its structure is early-type, that is with S\'ersic index $n > 2.5$.
\item It is compact: its size must be close to stellar mass--size relationship of early types at $z \sim 2$ (as indicated in Fig.~\ref{fig3}).
\end{enumerate}

\subsubsection{Results from the observations}

\begin{figure}
\center
\includegraphics{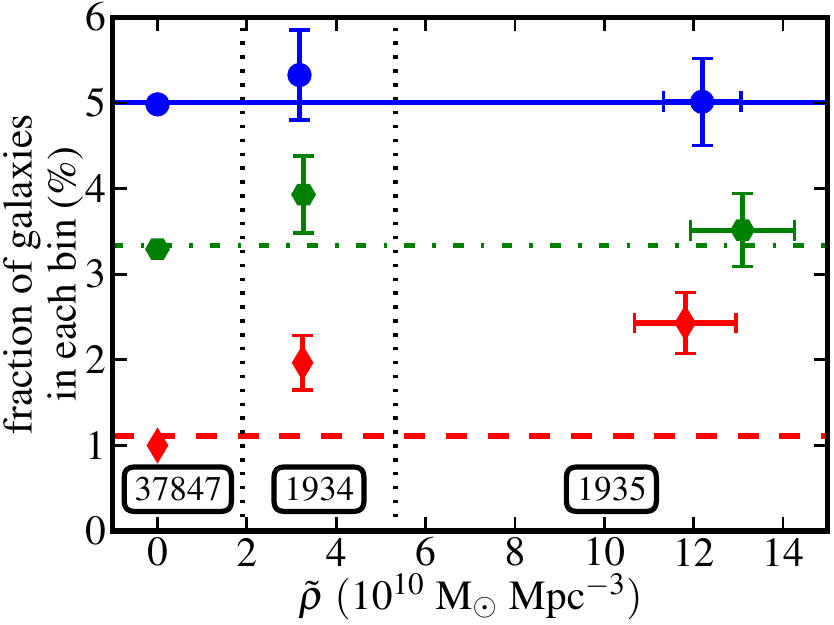}
\caption{\label{fig4}\, From the observational catalogue, fraction of normal-sized galaxies (blue circles), normal-sized early-type galaxies (green hexagons) and relics (red diamonds) in each environmental-density bin for the subsample. Horizontal blue solid, green dash--dotted and red dashed solid lines show the global fractions of galaxies (i.e. those computed neglecting any $\tilde{\rho}$ binning) which are normal-sized galaxies, normal-sized early-type galaxies and relics, respectively. Vertical black dotted lines show the limits used for binning in the $\tilde{\rho}$ axis, while the numbers within boxes indicate the number of galaxies of the whole sample which belong to each $\tilde{\rho}$ bin. The error bars with similar sizes to the symbol sizes were omitted for clarity. Figure from \cite{2016MNRAS.461..156P}.}
\end{figure}

\begin{itemize}
\item The fraction of relic galaxies is higher in environments with higher environmental densities (Fig.~\ref{fig4}).
\end{itemize}

\section{Takeaway messages}
\begin{itemize}
\item The fraction of relic galaxies is higher in environments with higher environmental densities.
\item Simulations show that relics tend to be located closer to the central parts of the clusters than the other massive galaxies.
\end{itemize}

%
%
\small  
%
\section*{Acknowledgments}   
%
This work has been supported by the Programa Nacional de Astronom\'{\i}a y Astrof\'{\i}sica of the Spanish Ministry of Economy and Competitiveness (under the grants AYA2009-11137, AYA2013-48226-C3-1-P and AYA2013-48226-C3-2-P) and the Generalitat Valenciana (under the grant PROMETEOII/2014/069).
%

%
\end{document}